\documentclass[showpacs,showkeys,prb,preprint,aps,preprintnumbers,amsmath,amssymb]{revtex4}
\usepackage{graphicx}
\usepackage{dcolumn}
\usepackage{bm}
\usepackage{multirow}
\begin{document}

\preprint{APS/123-QED}

\title{Anisotropic magnetization studies of R$_2$CoGa$_8$ \\(R = Gd, Tb, Dy, Ho, Er, Tm, Y and Lu) single crystals}
\author{Devang A. Joshi, R. Nagalakshmi, S. K. Dhar and  A. Thamizhavel}
\affiliation{Department of Condensed Matter Physics and Material
Sciences, Tata Institute of Fundamental Research, Homi Bhaba Road,
Colaba, Mumbai 400 005, India.}
\date{\today}

\begin{abstract}
Single crystals of R$_2$CoGa$_8$ series of compounds were grown, for the first time, by high temperature solution growth (flux) method. These compounds crystallize in a tetragonal crystal structure with the space group $P4/mmm$. It has been found that R$_2$CoGa$_8$ phase forms only with the heavier rare earths, starting from Gd with a relatively large $c/a$ ratio of $\approx$~2.6.  The resultant anisotropic magnetic properties of the compounds were investigated along the two principal crystallographic directions of the crystal viz., along [100] and [001]. The nonmagnetic compounds Y$_2$CoGa$_8$ and Lu$_2$CoGa$_8$ show diamagnetic behavior down to the lowest temperature (1.8~K) pointing out the non-magnetic nature of Co in these compounds and a relatively low density of electronic states at the Fermi level. Compounds with the magnetic rare earths order antiferromagnetically at temperatures lower than 30~K. The easy axis of magnetization for R$_2$CoGa$_8$ (R = Tb, Dy and Ho) is found to be along the [001] direction and it changes to [100] direction for Er$_2$CoGa$_8$ and Tm$_2$CoGa$_8$. The magnetization behavior is analyzed on the basis of crystalline electric field (CEF) model. The estimated crystal field parameters explains the magnetocrystalline anisotropy in this series of compounds.

\end{abstract}



\maketitle
\section {Introduction}
R$_{\rm n}$TX$_{\rm 3n+2}$ (R = rare earths, T = Co, Rh and Ir and X = In and Ga) form a family of compounds, mainly consisting of two groups with n = 1 and n = 2. Both the groups of compounds are structurally similar and exhibit a variety of interesting physical phenomena, which include heavy fermions, superconductivity and their coexistence, pressure induced superconductivity, magnetic ordering etc.  Compounds of n = 1 group have been investigated extensively compared to n = 2. The latter group of compounds (R$_2$TX$_8$) were first reported by Kalychak~\textit{et al}~\cite{Kalychak}, who reported the crystallographic details on these compounds. Later on, the interest in these compounds grew further due to the interesting behavior shown by the Ce compounds.  Ce$_2$RhIn$_8$ orders antiferromagnetically with a N\'{e}el temperature of 2.8~K  and it undergoes  pressure induced superconductivity at 2~K under a pressure of 2.3~GPa~\cite{Nicklas}.   Ce$_2$CoIn$_8$ is a Kondo lattice exhibiting heavy fermion superconductivity with a   $T_{\rm c}$ = 0.4~K~\cite{Chen, Devang} at ambient pressure, while Ce$_2$IrIn$_8$ is a heavy fermion paramagnet~\cite{Thompson}. 

The magnetic properties of polycrystalline R$_2$CoIn$_8$ compounds have been reported by Devang~\textit{et al}~\cite{Devang}. These compounds crystallize in the tetragonal structure with the space group \textit{P4/mmm}. Some of their interesting features are:  a crystal field split nonmagnetic doublet ground state of Pr$^{3+}$ ions in Pr$_2$CoIn$_8$,  field induced ferromagnetic behavior at low temperatures in Dy$_2$CoIn$_8$ and Ho$_2$CoIn$_8$ and an anomalously high magnetoresistance ($\sim$~2700~\%) at 2~K in Tb$_2$CoIn$_8$. Considering these interesting behaviors in the indium analogs, and to the best of our knowledge there are no reports on the corresponding gallium compounds, we decided to study the R$_2$CoGa$_8$ for various rare earths.  Here we report on our detailed structural and magnetization studies in this series of single crystals.

\section{Experiment}

Single crystals of R$_2$CoGa$_8$ (R = rare earths) compounds were grown by the flux method.  The starting materials used for the preparation of R$_2$CoGa$_8$ single crystals were high purity metals of rare-earths (99.95\%), Co (99.9\%) and Ga (99.999\%).  Owing to the low melting point of Ga, the single crystals were grown in Ga flux.  Considering the great affinity of Co and Ga atoms to form CoGa$_3$, we decided to make R$_2$Co button by arc melting and use it with excess Ga flux. From the binary phase diagram of R-Co we found that R$_2$Co phase does not exist, so the melt formed will have arbitrary phases which are assumed to remain unstable in presence of Ga within the required temperature range. From the previous studies~\cite{Chen}, it was observed that the R$_2$CoIn$_8$ crystals were grown in the temperature range between 750 and 450~$^\circ$C. We have also employed the same temperature range for the growth of R$_2$CoGa$_8$ compounds.  A button of R$_2$Co with excess of Ga (R:Co:Ga = 2:1:27) was taken in an alumina crucible and then sealed in an evacuated quartz ampoule. The ampoule was then heated up to 1050~$^\circ$C over a period of 24 hours and held at this temperature for 24 hours, so that the melt becomes homogeneous.  The furnace was then cooled very rapidly down to 750~$^\circ$C to avoid the formation of any unwanted phase. From 750 to 400~$^\circ$C the furnace was cooled down at the rate of 1~$^\circ$C/h, followed by a fast cooling to room temperature. The crystals were separated by centrifuging and as well treating them in hot water. The crystals obtained were platelets of size roughly 5~$\times~5~\times$~1~mm$^3$. In some of the cases the small platelets stick together to form a big crystal (roughly 7~$\times~5~\times$~ 4 mm$^3$). An energy dispersive X-ray analysis (EDAX) was performed on all the obtained single crystals to identify their phase. The EDAX results confirmed the crystals to be of the composition 2:1:8. To check for the phase purity,  powder x-ray diffraction pattern of all the compounds were recorded by powdering a few small pieces of single crystal. The R$_2$CoGa$_8$ phase forms only for heavier rare earths (Gd, Tb, Dy, Ho, Er, Tm and Lu, Y). Our attempts to make the compound with lighter rare earths failed. We did not attempt to make Yb and Eu compounds. It was found that in all the cases the crystallographic (001) plane was perpendicular to the flat plates of the crystal.  The crystals were oriented along the crystallographic axis [100] and [001] using Laue X-ray diffractometer, and cut along the principal directions for the purpose of magnetization measurements by spark erosion cutting machine. The magnetic measurements were performed using a superconducting quantum interference device (SQUID - Quantum Design) and vibrating sample magnetometer (VSM Oxford Instruments). 

\section{Results}
\subsection {Crystal structure}
\begin{figure}[b]
\includegraphics[width=0.5\textwidth]{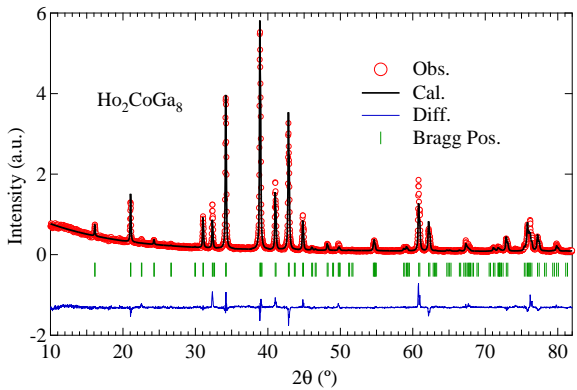}
\caption{\label{fig1}Powder X-ray diffraction pattern recorded for
crushed single crystals of Ho$_2$CoGa$_8$ at room temperature. The
solid line through the experimental data points is the Rietveld
refinement profile calculated for the tetragonal Ho$_2$CoGa$_8$.}
\end{figure}

\begin{figure}
\includegraphics[width=0.5\textwidth]{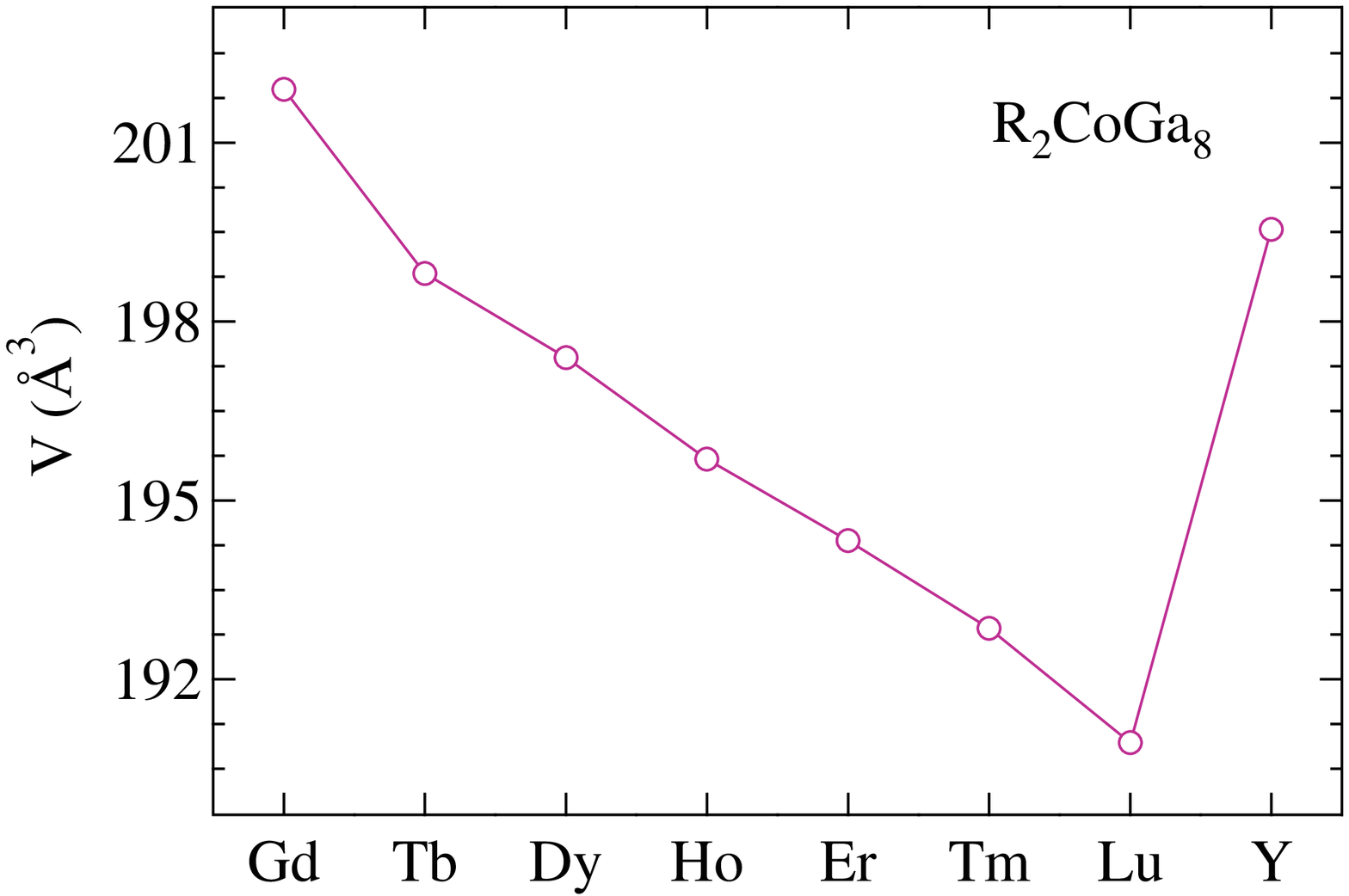}
\caption{\label{fig2} (Color online) Unit cell volume of  R$_2$CoGa$_8$ compounds  plotted against the corresponding rare earths.}
\end{figure}

\begin{figure}
\includegraphics[width=0.35\textwidth]{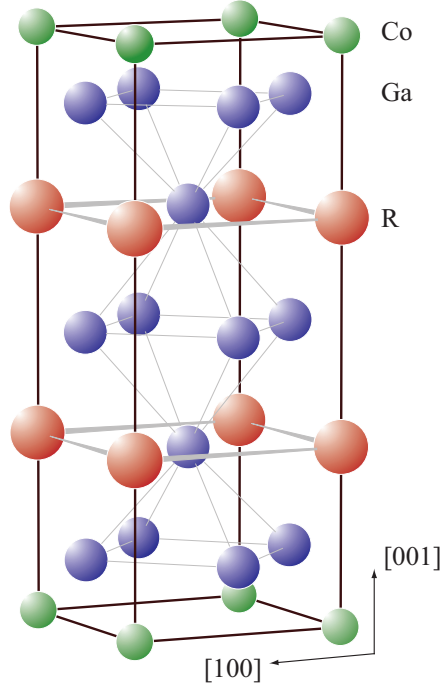}
\caption{\label{fig3} Tetragonal unit cell of R$_2$CoGa$_8$ compounds.}
\end{figure}

The R$_2$CoGa$_8$ series of compounds form in a tetragonal structure with a space group \textit{P4/mmm} (\# 123). One of the compounds of the series, Ho$_2$CoGa$_8$ is well cited for this particular structure. In order to confirm the phase homogeneity of the compound with proper lattice and crystallographic parameters, a Rietveld analysis of the observed X-ray pattern of all the compounds was done. The lattice parameters thus obtained are listed in Table~\ref{table1} and a representative Rietveld refined plot of Ho$_2$CoGa$_8$ is shown in Fig.~\ref{fig1}. The crystallographic parameters for each 
%
\begin{table}[!]
\centering
\begin{ruledtabular}
\begin{tabular}{ccccc}

Rare earth & \multicolumn{2}{c}{Lattice parameter} & Volume & $c/a$ \\
\cline{2-3} 
          &  $a$~(\r{A})  &  $b$~(\r{A})  &  (\r{A}$^3$)  & \\
\hline 
Gd & ~~~4.265~~~ & ~~~11.099~~~ & ~~~201.89~~~ & ~~~2.602~~~ \\ 
Tb & ~~~4.243~~~ & ~~~11.043~~~ & ~~~198.80~~~ & ~~~2.602~~~ \\ 
Dy & ~~~4.231~~~ & ~~~11.027~~~ & ~~~197.39~~~ & ~~~2.606~~~ \\ 
Ho & ~~~4.219~~~ & ~~~10.994~~~ & ~~~195.69~~~ & ~~~2.605~~~ \\
Er & ~~~4.210~~~ & ~~~10.964~~~ & ~~~194.32~~~ & ~~~2.604~~~ \\
Tm & ~~~4.199~~~ & ~~~10.938~~~ & ~~~192.85~~~ & ~~~2.604~~~ \\
Lu & ~~~4.181~~~ & ~~~10.903~~~ & ~~~190.93~~~ & ~~~2.607~~~ \\
Y  & ~~~4.249~~~ & ~~~11.053~~~ & ~~~199.55~~~ & ~~~2.601~~~ \\ 

\end{tabular}
\end{ruledtabular}
\caption{Lattice parameters, unit cell volume and $c/a$ ratio for the R$_2$CoGa$_8$ series of compounds}
\label{table1}
\end{table}
\begin{table}[!]
\centering
\begin{ruledtabular}
\begin{tabular}{c|c|c|c|c|c|c}

Atom & Site  & x & y & z & U$_{\rm eq}$(\AA$^2)$ & Occ.\\ 
& symmetry & & & & & \\
\hline

Ho  & 2g & ~~~0~~~ & ~~~0~~~   & ~~~0.306~~~   & ~~~0.588~~~ & ~~~2~~~ \\ 
Co  & 1a & ~~~0~~~ & ~~~0~~~  & ~~~0~~~       & ~~~1.186~~~ & ~~~1~~~ \\ 
Ga1 & 2e & ~~~0~~~ & ~~~0.5~~~ & ~~~0.5~~~     & ~~~1.771~~~ & ~~~2~~~ \\ 
Ga2 & 2h & ~0.5~ & ~~~0.5~~~ & ~~~0.295~~~     & ~~~0.40~~~  & ~~~2~~~ \\
Ga3 & 4i & ~~0~~ & ~~~0.5~~~ & ~~~0.114~~~   & ~~~0.074~~~ & ~~~4~~~ \\ 

\end{tabular}
\end{ruledtabular}
\caption{Refined crystallographic parameters for Ho$_2$CoGa$_8$}
\label{table2}
\end{table}
of the constituent atoms (at various crystallographic sites) in Ho$_2$CoGa$_8$ are presented in Table~\ref{table2}. Both the lattice parameters  $a$ and $c$ are smaller than the corresponding In compounds~\cite{Devang}. One of the possible reasons is due to the smaller metallic radii of Ga ($\approx$~1.3~\AA) compared to that of In ($\approx$~1.6~\AA). The unit cell volumes of the R$_2$CoGa$_8$ compounds are plotted against their corresponding rare earths in Fig~\ref{fig2}. The unit cell volume and both the lattice parameters 
decrease as we move from Gd to Lu. This is attributed to the well known lanthanide contraction. The  \textit{(c/a)}  ratio remains constant ($\approx$~2.6) for all the compounds but it is slightly less than that of the corresponding In compounds ($\approx$~2.63). The large \textit{(c/a)} ratio indicates the significant structural anisotropy in these compounds. The crystal structure of R$_2$CoGa$_8$ is shown in Fig.~\ref{fig3}.
The central portion of the unit cell between the rare earth planes (along the $c$ axis) and including them is similar to  the unit cell of RGa$_3$ phase. In fact, the lattice constant $a$ of R$_2$CoGa$_8$ phase is approximately equal to that of the RGa$_3$ (cubic) phase~\cite{Cirafici} and the remaining structure above and below it forms the CoGa$_2$ layers.   The unit cell of R$_2$CoGa$_8$ may be viewed as formed by the stacking of  RGa$_3$ units and CoGa$_2$ layer alternately along the \textit{c} axis.  It is imperative to mention here one similarity between RGa$_3$ and R$_2$CoGa$_8$ compounds; viz., in both cases the compounds form only with heavy rare earths. Gd is on the borderline of stability which forms the R$_2$CoGa$_8$ phase but not the RGa$_3$ phase~\cite{Cirafici}.  A similar comparison is also possible for indides (R$_2$CoIn$_8$~\cite{Kalychak2} and RIn$_3$~\cite{Buschow}), where only Lu and La are on the borderline of stability and do not form R$_2$CoIn$_8$ phase, where as RIn$_3$ forms for the entire rare earth series. Hence the non formation of RGa$_3$ phase with lighter rare earths is one of the possible reasons for the non formation of the corresponding R$_2$CoGa$_8$ compounds.  Therefore, RX$_3$ (X = In and Ga) can plausibly be considered as one of the basic building blocks for the formation of R$_2$CoX$_8$ (X = In and Ga) compounds.

\section{Magnetic Properties}

\subsection{Y$_2$CoGa$_8$, Lu$_2$CoGa$_8$}

\begin{figure}
\includegraphics[width=0.5\textwidth]{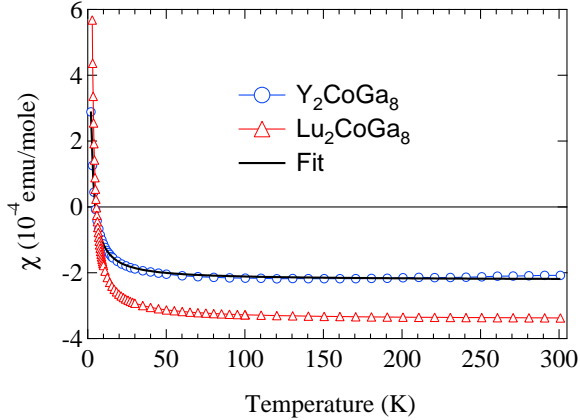}
\caption{\label{fig4}(Color online) Temperature dependence of the magnetic susceptibility in Y$_2$CoGa$_8$ and Lu$_2$CoGa$_8$.  The solid line through the data points implies the modified Curie-Weiss fitting. }
\end{figure}

These two compounds with the non-magnetic Y and Lu respectively, show diamagnetic behavior. The susceptibility of both the compounds is shown in Fig.~\ref{fig4}.  It is negative and nearly temperature independent at high temperatures and exhibits a weak upturn at low 
temperatures crossing into the positive region below 10~K.  Assuming a magnetic moment of 1~$\mu_{\rm B}$/impurity, a good fit of the modified Curie-Weiss expression 
\begin{equation}
\label{eqn1}
\chi = \chi_0 + \frac{C}{T-\theta_{\rm p}},
\end{equation}
to the data lead to  paramagnetic impurity ion concentration of a few ppm and $\theta_{\rm p}$ is nearly zero. The value of $\chi_0$ is found to be -2.08~$\times$~10$^{-4}$~emu/mol and  -3.13~$\times$~10$^{-4}$~emu/mol for Y$_2$CoGa$_8$ and Lu$_2$CoGa$_8$, respectively.  These data show conclusively the nature of non-magnetic Co atoms in this series of compounds.  

The diamagnetic behavior of Y$_2$CoGa$_8$ and Lu$_2$CoGa$_8$ is in contrast with the Pauli-paramagnetic behavior shown by the non magnetic indide Y$_2$CoIn$_8$ and indicates a low density of electronic density of states at the Fermi level in the gallides, such that the diamagnetic contribution due to the filled electronic shells exceeds the Pauli paramagnetic contribution from the conduction electrons.   Indeed, the co-efficient of the electronic heat capacity ($\gamma$),  which is proportional to  the density of states at the Fermi level, of Y$_2$CoGa$_8$ is (2~mJ/K$^2\cdot$mol, comparably less than that of Y$_2$CoIn$_8$~\cite{Devang} (13~mJ/K$^2\cdot$mol).  If the density of states at the Fermi level decreases on replacing In by Ga in these R$_2$CoX$_8$ (X = In and Ga) series of compounds, it should result in a weaker conduction electron mediated RKKY magnetic interactions between the rare earth ions in R$_2$CoGa$_8$ compounds.  This may be one of the reasons that the N\'{e}el temperatures of R$_2$CoGa$_8$ series of compounds are lower than those of the corresponding indides R$_2$CoIn$_8$.

\subsection{Gd$_2$CoGa$_8$}

\begin{figure}
\includegraphics[width=0.5\textwidth]{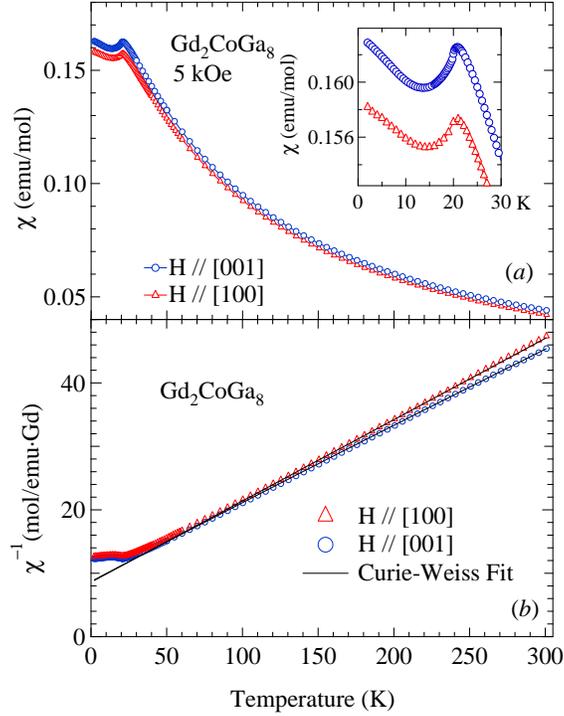}
\caption{\label{fig5} (Color online) (a) Magnetic susceptibility of Gd$_2$CoGa$_8$, inset shows the magnified view of low temperature susceptibility, (b) inverse magnetic susceptibility with a modified Curie-Weiss fit.}
\end{figure}

\begin{figure}[!]
\includegraphics[width=0.5\textwidth]{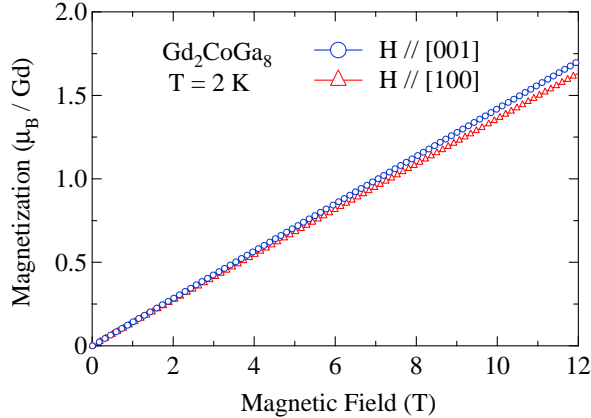}
\caption{\label{fig6} (Color online) Magnetic isotherm of Gd$_2$CoGa$_8$ at 2~K with the field along [100] and [001] directions.}
\end{figure}

We next describe the magnetization of  Gd$_2$CoGa$_8$, as Gd is a $S$ state ion in which the crystal electric field induced anisotropy is zero in the first order.  The susceptibility of Gd$_2$CoGa$_8$ along [100] and [001] directions in an  applied magnetic field of 5~kOe is shown in Fig.~\ref{fig5}(a). The low 
temperature part is shown as an inset of Fig.~\ref{fig5}(a). The susceptibility shows a peak due to antiferromagnetic transition at $T_{\rm N}$ = 20~K, followed by an upturn below $\approx$~15~K.  Quantitatively similar behavior is seen with the field applied in [100] and [001] directions, respectively.  The minor difference in the susceptibility along the two axes may be due to the second order anisotropy arising from the dipole-dipole interaction. The linear behavior of the magnetic isotherms at 2~K (Fig.~\ref{fig6}) for 
both the axes further corroborates the antiferromagnetic nature  of the magnetically ordered state. In the paramagnetic state the susceptibility was fitted to the modified Curie-Weiss law as shown in Fig~\ref{fig5}(b). The obtained effective magnetic moments presented in  Table~\ref{table3} are close to that of the theoretically expected one. The paramagnetic Curie temperatures $\theta_{\rm p}$ are -69~K and -67~K respectively, for [100] and [001] directions. The relatively high value along both the directions indicate strong antiferromagnetic interaction among the Gd$^{3+}$ moments. The upturn in the susceptibility at low temperatures is often attributed to a canting of antiferromagnetically aligned moments.  A similar behavior has earlier been seen in GdCo$_2$Si$_2$~\cite{Mallik}, which was later shown to be due to a non-collinear amplitude modulated structure~\cite{Rotter}.  So we assume that a similar or some complicated stable magnetic structure is present in Gd$_2$CoGa$_8$.

\subsection{R$_2$CoGa$_8$ (R = Tb, Dy and Ho)}

\begin{figure}[!]
\includegraphics[width=0.5\textwidth]{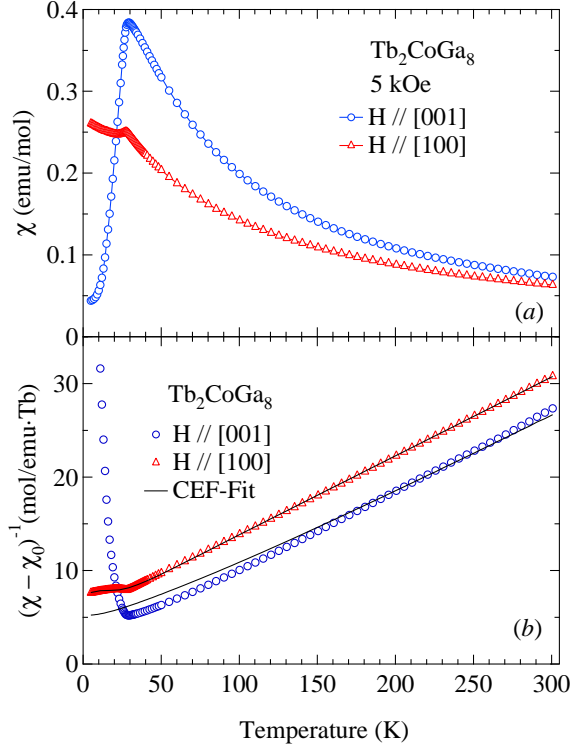}
\caption{\label{fig7} (Color online) (a) Magnetic susceptibility of Tb$_2$CoGa$_8$,  (b) inverse magnetic susceptibility; solid lines through the data point indicate the CEF fit.}
\end{figure}

In these three compounds the easy-axis of magnetization lies along the [001]  direction. The data clearly show the anisotropic behavior of the magnetization in both the paramagnetic and anitferromagnetically ordered states arising due to the influence of crystal electric fields (CEF) on the Hund's rule derived ground states of the free $R^{3+}$ ions.  Figure~\ref{fig7}(a) shows the susceptibility of Tb$_2$CoGa$_8$ from 1.8 to 300~K in a magnetic field of 5~kOe along the two crystallographic directions ([100] and [001]). The data show an antiferromagnetic transition at T$_{\rm N}$~=~28~K. The 
anisotropic behavior of the susceptibility below the N\'{e}el temperature shows [001] direction as the easy axis of magnetization. In contrast, the susceptibility along [100] remains below that of [001] direction in the entire temperature range followed by a knee at the ordering temperature. The Curie-Weiss fits of the inverse susceptibility in the paramagnetic state gives $\mu_{\rm eff}$ and $\theta_{\rm p}$ as 9.55~$\mu_{\rm B}$/Tb and 9.66~$\mu_{\rm B}$/Tb and -58~K and -16~K for field parallel to [100] and [001] directions, respectively. The obtained effective moment for both the axes is close to that of the theoretical value (9.72~$\mu_{\rm B}$/Tb). The polycrystalline average of $\theta_{\rm p}$ is -45.3~K, indicating the presence of strong antiferromagnetic interactions in the compound.

\begin{figure}[!]
\includegraphics[width=0.5\textwidth]{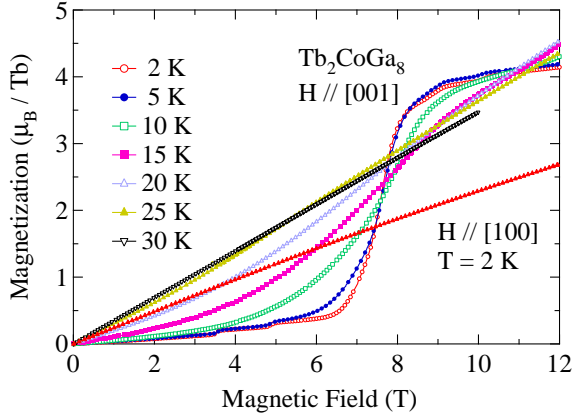}
\caption{\label{fig8} (Color online) Magnetic isotherms of Tb$_2$CoGa$_8$ at 2~K for the field along [100] and at various temperatures for [001] direction.}
\end{figure}

The anisotropic magnetic behavior is further corroborated by the magnetic isotherms (measured at various temperatures along the [001] direction) of the compound with field applied along the two crystallographic axes, respectively which is shown in Fig.~\ref{fig8}(a). 
The magnetization at T~=~2~K undergoes multiple metamagnetic transitions at 3.5, 4.8, 6.6, 9 and 10~T. The magnetic transition at 6.6~T is a predominant one whereas the others represent minor reorientation of the moments with field. The saturation magnetization obtained at 12~T at 2~K is about 4.2 $\mu_{\rm B}$/Tb, which is less than half of the saturation moment of Tb$^{3+}$ ions (9~$\mu_{\rm B}$/Tb). Further high magnetic field is required to obtain the full saturation value of the Tb moments.   As the temperature is increased the sharpness of the metamagnetic transition decreases and it shifts towards lower fields due to the  extra thermal energy available for the reorientation of the moments.  At 30~K the magnetization is linear indicating the paramagnetic state of the compound. The magnetic isotherm at 2~K with the field along the hard direction [100] (represented by triangles in Fig.~\ref{fig8}(a)) is a straight line with a magnetization value of 2.6~$\mu_{\rm B}$/Tb at 12~T.

\begin{figure}[!]
\includegraphics[width=0.5\textwidth]{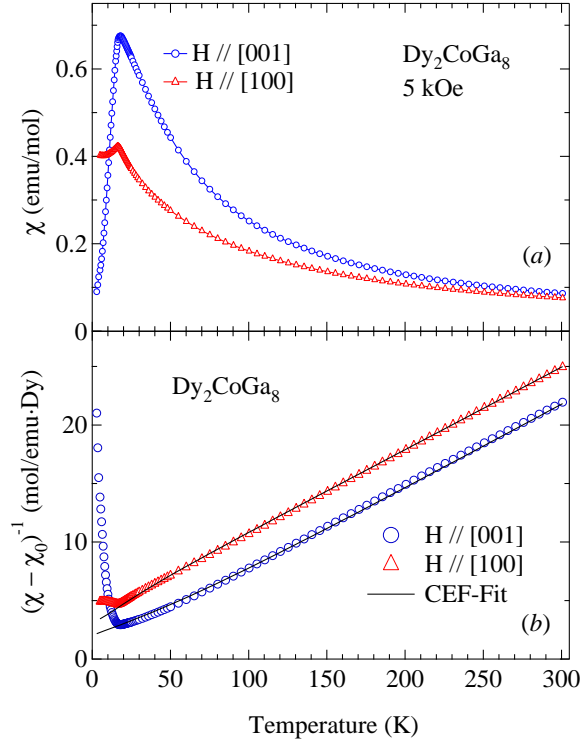}
\caption{\label{fig9} (Color online) (a) Magnetic susceptibility of Dy$_2$CoGa$_8$,  (b) inverse magnetic susceptibility; solid lines through the data point indicate the CEF fit.}
\end{figure}

\begin{figure}[t]
\includegraphics[width=0.5\textwidth]{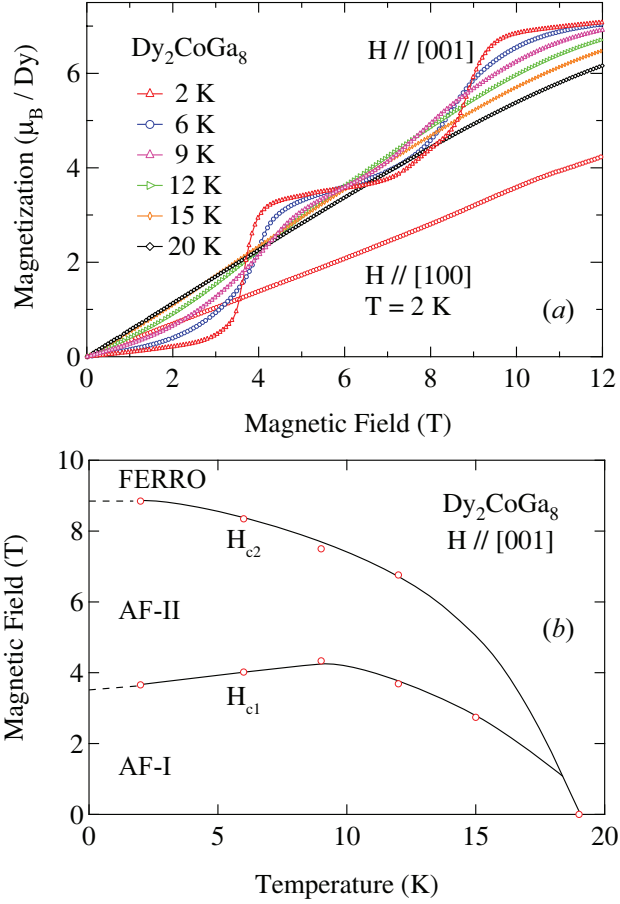}
\caption{\label{fig10} (Color online) Magnetic isotherms of Tb$_2$CoGa$_8$ at 2~K for the field along [100] and at various temperatures for [001] direction and (b) the magnetic phase diagram of Tb$_2$CoGa$_8$.}
\end{figure}

The thermal variation of the  magnetic susceptibility of Dy$_2$CoGa$_8$ is similar to that of Tb$_2$CoGa$_8$. An antiferromagnetic transition occurs at $T_{\rm N}~$~=18~K as shown in Fig.~\ref{fig9}(a). 
In the paramagnetic state the inverse susceptibility was fitted to Curie-Weiss law with $\mu_{\rm eff}$~=~10.4~$\mu_{\rm B}$/Dy and 10.5~$\mu_{\rm B}$/Dy and $\theta_{\rm p}$ = -45~K and -6~K along the [100] and [001] axis, respectively. The effective moments are close to the theoretical value (10.63~$\mu_{\rm B}$) and the polycrystalline average of $\theta_{\rm p}$ is -32~K indicating an antiferromagnetic interaction. The magnetic isotherms with field along the two crystallographic directions are shown in Fig.~\ref{fig10}(a). The temperature variation of magnetic isotherms was measured only along the easy axis [001]. The magnetization along the [001]  at 2~K undergoes two metamagnetic transitions at $H_{\rm c1}$ = 3.4~T and $H_{\rm c2}$ = 8.6~T. 
After the second metamagnetic transition the magnetization reaches about 7.1~$\mu_{\rm B}$/Dy at 12~T. This value is less than the ideal saturation value of 10~$\mu_{\rm B}$/Dy for Dy$^{3+}$ ion.  A hysteresis was observed (not shown in the figure) between the two metamagnetic transitions, which may be due to the anisotropic behavior of the reoriented moments. The temperature variation of the magnetic isotherm is similar to that of Tb$_2$CoGa$_8$, namely the decrease in the sharpness and shift towards lower magnetic fields of the metamagnetic transition with temperature. In Tb$_2$CoGa$_8$ the magnetization at 12~T was found to increase initially with temperature and then decreases near the N\'{e}el temperature of the compound, whereas in Dy$_2$CoGa$_8$ it decreases continuously with temperature.  This effect is attributed to the strong antiferromagnetic coupling of the Tb$^{3+}$ moments compared to that of Dy$^{3+}$ moments.  This is also evident from the polycrystalline average $\theta_{\rm p}$ of both the compounds.   Because of the strong coupling the thermal energy acts as a helping hand for the reorientation of the moments, whereas in Dy$_2$CoGa$_8$ the field energy is sufficient to break the antiferromagnetic coupling. From the differential plots of the isothermal magnetization curves (not shown here), we have constructed the magnetic phase diagram as depicted in Fig.~\ref{fig10}(b). $H_{\rm c1}$ at first increases with the increase in temperature and then decreases above 10~K and finally vanishes for temperature above 15~K , while the $H_{\rm c2}$ decreases continuously with the increase in the temperature.  At low temperatures and for fields less than 3.4~T, the systems is in a purely antiferromagnetic state as indicated by (AF-I) in Fig.~\ref{fig10}(b) and then undergoes a complex magnetic structure (AF-II) for fields between $H_{\rm c1}$ and $H_{\rm c2}$ and finally enters into the field induced ferromagnetic state.  

\begin{figure}[t]
\includegraphics[width=0.5\textwidth]{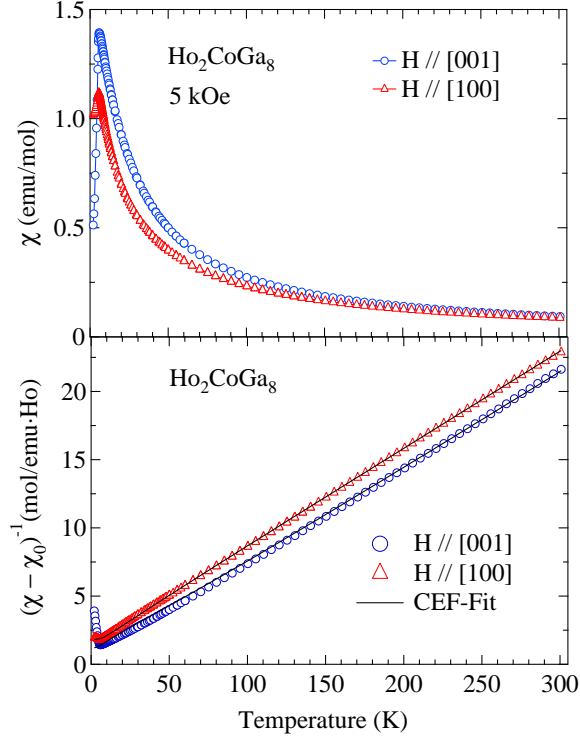}
\caption{\label{fig11} (Color online) (a) Magnetic susceptibility of Ho$_2$CoGa$_8$,  (b) inverse magnetic susceptibility; solid lines through the data point indicate the CEF fit.}
\end{figure}

\begin{figure}[t]
\includegraphics[width=0.5\textwidth]{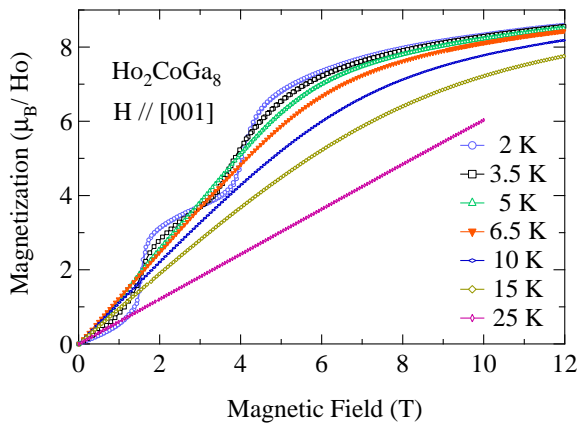}
\caption{\label{fig12} (Color online) Magnetic isotherms of Ho$_2$CoGa$_8$ at 2~K for the field along [100] and at various temperatures for [001] direction.}
\end{figure}

The magnetic susceptibility of Ho$_2$CoGa$_8$ in an applied magnetic field of 5~kOe along [100] and [001] axes is shown in Fig.~\ref{fig11}(a).  The compound orders antiferromagnetically at $T_{\rm N}$ = 6~K.
Overall, the susceptibility along the [001] and [100] direction shows a similar behavior as observed in  Tb$_2$CoGa$_8$ and Dy$_2$CoGa$_8$ analogs. This indicates a less anisotropic behavior of Ho$_2$CoGa$_8$ compared to the former two compounds. The Curie-Weiss fit of the inverse susceptibility in the paramagnetic state of the compound gives $\mu_{\rm eff}$ = 10.48~$\mu_{\rm B}$/Ho and 10.6~$\mu_{\rm B}$/Ho and $\theta_{\rm p}$ = -18.6~K and -1.5~K along the [100] and [001] direction respectively. The effective moments are close to the theoretically expected value (10.6~$\mu_{\rm B}$/Ho) and the polycrystalline average of $\theta_{\rm P}$ is -12.9~K, in accordance with the  antiferromagnetic transition in the compound. A lower absolute value of $\theta_{\rm p}$ compared to those of compounds with R = Gd, Tb and Dy, indicates weaker interaction among the Ho$^{3+}$ moments. This is evident in the magnetic isotherm of the compound along the easy axis of magnetization (as shown in Fig.~\ref{fig12}). The  metamagnetic transition occurs at lower fields compared to the former compounds. The magnetic isotherm at 2~K (Fig.~\ref{fig12}) along the [001] axis undergoes two metamagnetic transitions at $H_{\rm c1}$~=~1.55~T and at $H_{\rm c2}$~=~4.0~T. Both of them induce large 
changes in the magnetization. The magnetization at 12~T and 2~K is  8.2~$\mu_{\rm B}$/Ho. This is close to the saturation moment of 10~$\mu_{\rm B}$ for Ho$^{3+}$ moments. Thus the second metamagnetic transition drives the compound to a field induced ferromagnetic state. With the increase in the temperature the metamagnetic transitions broaden and shifts towards lower fields and their sharpness decreases. Above the ordering temperature of the compound (6~K), the magnetic isotherms at 10 K and 15 K tend towards saturation at high fields and the moment at 12~T is above 8.2~$\mu_{\rm B}$/Ho. At 25~K the magnetic isotherm is a straight line as expected for a paramagnetic state. The magnetization with field along the hard direction [100] is not a straight line as observed for other compounds, but the moment tends towards saturation with field; $\approx$~7~$\mu_{\rm B}$/Ho at 12~T.
\begin{figure}[t]
\includegraphics[width=0.5\textwidth]{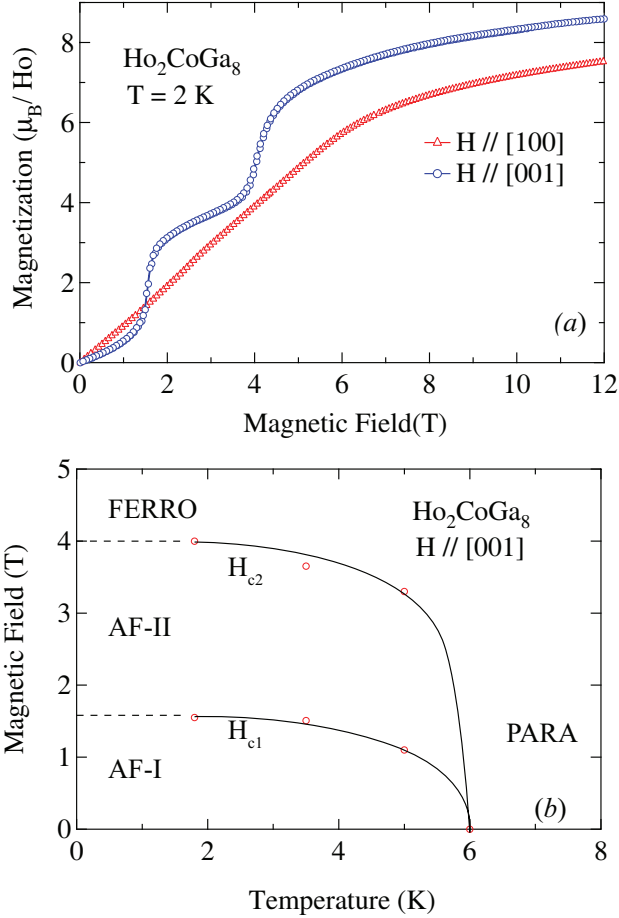}
\caption{\label{fig13} (Color online)(a) Magnetic isotherm of Ho$_2$CoGa$_8$ at 2~K for the field along [100] and [001] directions, (b) the magnetic phase diagram of Ho$_2$CoGa$_8$.  }
\end{figure}
At high fields (above 8~T), the behavior of the compound is similar in both the directions (except for the moment is little less along the hard direction). This is because the field energy is sufficient to overcome the anisotropic energy barrier existing within the compound. This also supports the early explanation of less anisotropy of Ho$_2$CoGa$_8$ compared to that of Tb$_2$CoGa$_8$ and Dy$_2$CoGa$_8$. From the differential plots of the isothermal magnetization along [001] direction we have constructed the magnetic phase diagram as shown in Fig~\ref{fig13}(b).  It is obvious from the figure that both the metamagnetic transitions shift towards lower fields with the increase in temperature and finally merge with each other at the ordering temperature.  

\subsection{R$_2$CoGa$_8$ (R = Er and Tm)}

Though these compounds also order antiferromagnetically, the easy axis of magnetization is now along the [100] direction.  Er$_2$CoGa$_8$ and Tm$_2$CoGa$_8$ order antiferromagnetically at 2 and 3~K, respectively.  The change in the direction of  the easy axis of magnetization follows the general trend observed in a number of tetragonal rare-earth series of compounds RRh$_4$B$_4$\cite{Dunlap}, RAgSb$_2$~\cite{Myers} and RRhIn$_5$~\cite{Hieu}. 
\begin{figure}[!]
\includegraphics[width=0.5\textwidth]{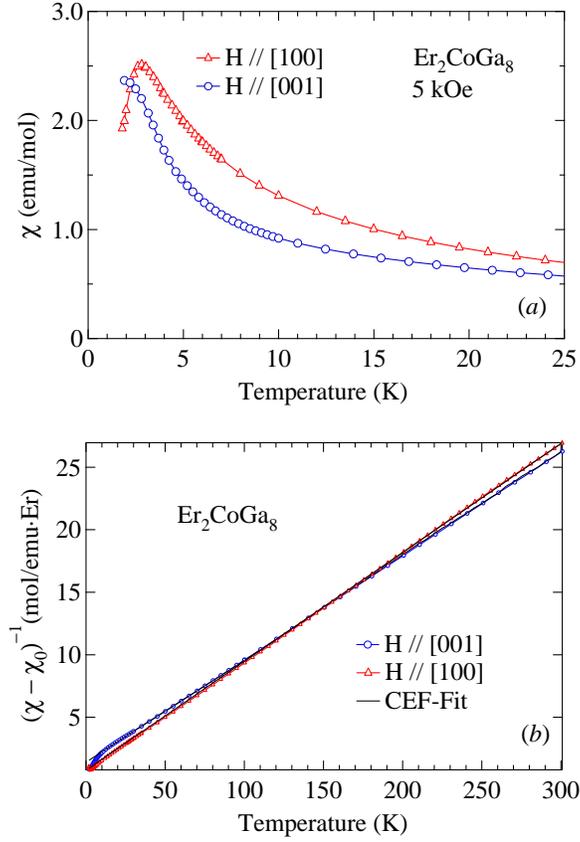}
\caption{\label{fig14} (Color online) (a) Magnetic susceptibility of Er$_2$CoGa$_8$ for the temperature range 1.8 to 25~K,  (b) inverse magnetic susceptibility; solid lines through the data point indicate the CEF fit.}
\end{figure}
The obtained value of  $\mu_{\rm eff}$ and $\theta_{\rm p}$ are presented in Table~\ref{table3}. The effective moment for both the compounds along both the crystallographic directions is close to their theoretically expected value (9.59~$\mu_{\rm B}$/Er and 7.57~$\mu_{\rm B}$/Tm). The polycrystalline average of the paramagnetic Curie temperature for Er$_2$CoGa$_8$ and Tm$_2$CoGa$_8$ are -7.13~K and -4.2~K respectively. Both of them are negative and are lower compared to the other R$_2$CoGa$_8$ compounds, indicating a gradual  weakening of the antiferromagnetic interactions among the rare earth moments, in this series.

\begin{figure}[!]
\includegraphics[width=0.5\textwidth]{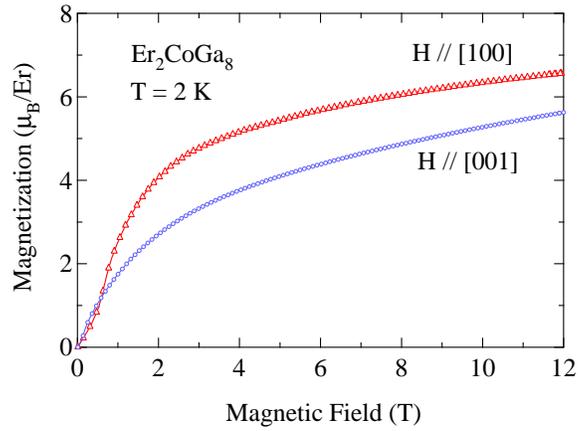}
\caption{\label{fig15} (Color online) Magnetic isotherm of Er$_2$CoGa$_8$ at 2~K for the field along [100] and [001] directions.}
\end{figure}

The susceptibility of Er$_2$CoGa$_8$ along both the directions is shown in Fig.~\ref{fig14}a. The susceptibility along the [100] direction shows a peak at the antiferromagnetic transition temperature, where as that along the [001] direction does not show any ordering down to 1.8~K.  The inverse susceptibility is plotted in Fig.~\ref{fig14}(b). Both the plots are close to each other indicating a weak anisotropy of the compound. In addition, there is a crossover of the susceptibility at about 164~K, indicating a change in easy axis of magnetization with temperature. Similar behavior has been observed in Er$_2$PdSi$_3$~\cite{Frontzek}. Such a behavior may possibly arise due to the magnetic behavior of the compound lying on the border of the anisotropic crossover along the crystallographic axis. This is also supported by the crystal field calculation on the compound discussed in  the next section. The magnetic isotherm along both the direction is shown in Fig.~\ref{fig15}. The magnetization along the [100] direction undergoes a metamagnetic transition at 0.25~T and tends to saturate at higher fields. The moment at 1.8~K and 12~T is 9~$\mu_{\rm B}$/Er.   With increase in the temperature (at 4~K and 10~K, not shown in the figure) the metamagnetic transition vanishes but the behavior of the magnetization remains the same with the moment little less then that at 2~K.  The magnetization behavior along the [001] direction is almost similar to that of [100] direction.    For field parallel to [100] the magnetization reaches a moment of  5.6~$\mu_{\rm B}$/Er at 1.8~K and 12~T.

\begin{figure}
\includegraphics[width=0.5\textwidth]{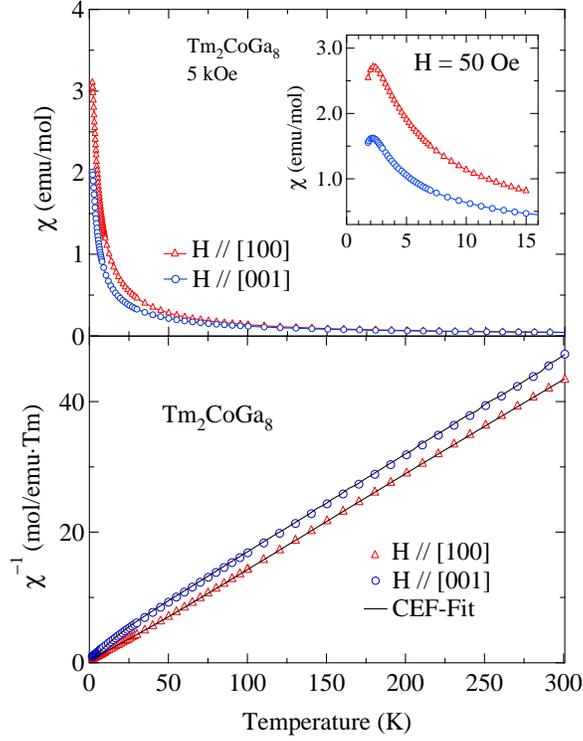}
\caption{\label{fig16} (Color online)(a) Magnetic susceptibility of Tm$_2$CoGa$_8$, inset shows the magnified view of low temperature susceptibility, (b) inverse magnetic susceptibility;  solid lines through the data point indicate the CEF fit.}
\end{figure}

\begin{figure}
\includegraphics[width=0.5\textwidth]{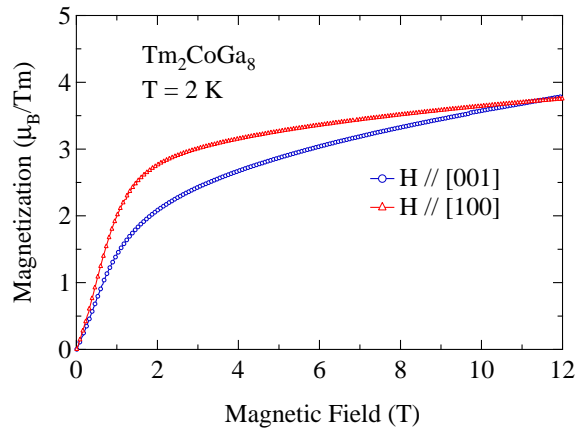}
\caption{\label{fig17} (Color online) Magnetic isotherm of Tm$_2$CoGa$_8$ at 2~K for the field along [100] and [001] directions.}
\end{figure}

In case of Tm$_2$CoGa$_8$ the susceptibility along both the crystallographic direction show antiferromagnetic behavior at 50~Oe (inset of Fig.~\ref{fig16}a). When the field is increased to 5~kOe, antiferromagnetic transition along both the axis vanishes. This indicates a weak interaction among the moments such that the antiferromagnetic peak shifts with small applied fields and vanishes even at 5~kOe. The magnetic isotherm at 1.8~K for both the crystallographic axis is shown in Fig.~\ref{fig17}. The 
behavior of magnetization along both the axis is qualitatively similar.  But the magnetization along the [100] direction is higher than that of the [001] direction. At approximately 11.2~T there is a cross over with the magnetization along the [001] axis exceeding the magnetization along the [100] axis.   High field magnetization measurements are necessary to probe this behavior.  

\section{Discussion}

An interesting magnetic behavior is exhibited by the R$_2$CoGa$_8$ series of compounds. The nonmagnetic compounds Y$_2$CoGa$_8$ and Lu$_2$CoGa$_8$ show diamagnetic behavior. Compounds with magnetic rare earths (R = Gd, Tb, Dy, Ho, Er and Tm) order antiferromagnetically at low temperatures. The N\'{e}el temperatures are listed in Table~\ref{table3}. For comparison the N\'{e}el temperatures of the corresponding indides are also listed.  The 
\begin{table*}[!]
\centering
\begin{ruledtabular}
\begin{tabular}{ccccccccc}
& R$_2$CoIn$_8$ & \multicolumn{7}{c}{R$_2$CoGa$_8$} \\
\cline{3-9}\\
&$T_{\rm N}$ & $T_{\rm N}$ &  \multicolumn{2}{c}{$\mu_{\rm eff}~(\mu_{\rm B}$/R)} & \multicolumn{2}{c}{$\theta_{\rm p}$~(K)} & \multicolumn{2}{c}{$\chi_{0}$~({emu/mol})} \\
\cline{4-9} \\

R &   (K)  & (K) & [100] & [001] & [100] & [001] & [100] & [001] \\
\hline \\
Gd    &    33.5    & 20    & 7.94    & 7.9    & -69   &  -67   & 7.2~$\times$10$^{-4}$   & 3.4~$\times$10$^{-4}$\\

Tb    &    30    & 28    & 9.55    & 9.66    & -58   &  -16   &  0   &  0\\

Dy    &    17.4    & 18    & 10.53    & 10.47    & -45   &  -6   & 9.0~$\times$10$^{-4}$   & 1.2~$\times$10$^{-3}$\\

Ho    &    7.6    & 6    & 10.5    & 10.6    & -18.6   &  -1.5   & 8.0~$\times$10$^{-4}$   & 3.0~$\times$10$^{-4}$\\

Er    &    N.A    & 3    & 9.5    & 9.59    & -5.2   &  -11.2   & 0   & 0\\

Tm    &    N.A    & 2    & 7.57    & 7.35    & -1.7   &  -12.6   & 1.7~$\times$10$^{-3}$   & 2.5~$\times$10$^{-4}$\\

Lu    &    N.A    & Dia  &  - & - & - & - & - & - \\

Y     &    P.P    & Dia  &  - & - & - & - & - & -\\

\end{tabular}
\end{ruledtabular}
N.A :  Data not available, P.P :  Pauli paramagnetic, Dia:  Diamagnetic
\caption{Comparison of N\'{e}el temperatures of R$_2$CoIn$_8$ with R$_2$CoGa$_8$ compounds.  Paramagnetic Curie temperature ($\theta_{\rm p}$), effective magnetic moment ($\mu_{\rm eff}$) and $\chi_{0}$ for R$_2$CoGa$_8$ compounds.}
\label{table3}
\end{table*}
\begin{figure}[!]
\includegraphics[width=0.5\textwidth]{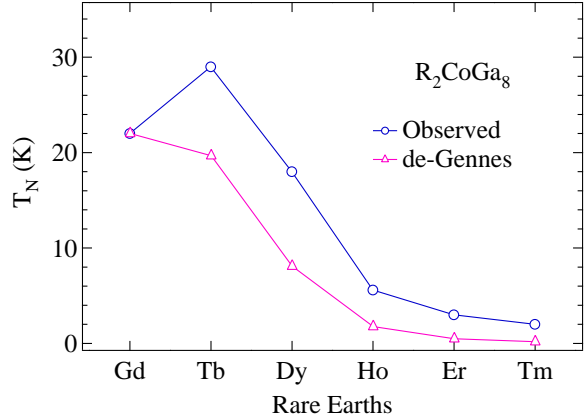}
\caption{\label{fig18} (Color online) N\'{e}el temperature of the R$_2$CoGa$_8$ compounds compared with that expected from de-Gennes scaling.  The lines joining the data points are the guide to eyes.}
\end{figure}
transition temperature is less compared to that of the corresponding polycrystalline indium analogs R$_2$CoIn$_8$. According to the de-Gennes scaling, in the mean field approximation the magnetic ordering temperatures $T_{\rm M}$ of  the isostructural members of a rare earth series of compounds are proportional to $(g_{\rm J} - 1)^2 J (J+1)$, where $g_{\rm J}$ is the Land\'{e} $g$ factor and $J$ is the total angular momentum. The N\'{e}el temperature of the R$_2$CoGa$_8$ series of compound is plotted in Fig.~\ref{fig18} along with their de-Gennes expected 
values, normalized to $T_{\rm N}$ of Gd$_2$CoGa$_8$. A fairly large deviation of T$_ {\rm N}$ of the Tb and Dy compounds from the de-Gennes expected scaled value is noticeable.  The ordering temperature of Tb$_2$CoGa$_8$ is even higher than that of the corresponding Gd compound. A similar behavior was also observed for RRh$_4$B$_4$~\cite{Dunlap} compounds and Noakes~\textit{et al.}~\cite{Noakes} showed it to be due to the CEF effects.  The magnetic susceptibility data presented above therefore provide a good opportunity to attempt a crystal electric field analysis on this series of compounds.  From  the estimated crystal field parameters we found that the enhancement in the ordering temperature of Tb and Dy can be attributed  to CEF effects, as described later in this section.

The rare-earth atom in R$_2$CoGa$_8$ occupies the  $2g$ Wyckoff's position with a tetragonal  $C_{4v}$ point symmetry.  The CEF Hamiltonian for a tetragonal symmetry is given by, 
\begin{equation}
\label{eqn2}
\mathcal{H}_{\rm CEF} = B_{2}^{0}O_{2}^{0} + B_{4}^{0}O_{4}^{0} + B_{4}^{4}O_{4}^{4} + B_{6}^{0}O_{6}^{0} + B_{6}^{4}O_{6}^{4},
\end{equation}
where $B_{\ell}^{m}$ and $O_{\ell}^{m}$ are the CEF parameters and the Stevens operators, respectively~\cite{Stevens,Hutchings}. The
CEF susceptibility is defined as
\begin{widetext}
\begin{equation}
\label{eqn3}
\chi_{{\rm CEF}i} = N(g_{J}\mu_{\rm B})^2 \frac{1}{Z}
\left(\sum_{m \neq n} \mid \langle m \mid J_{i} \mid n \rangle
\mid^{2} \frac{1-e^{-\beta \Delta_{m,n}}}{\Delta_{m,n}}e^{-\beta
E_{n}} + \sum_{n} \mid \langle n \mid J_{i} \mid n \rangle
\mid^{2} \beta e^{-\beta E_{n}} \right),
\end{equation}
\end{widetext}
where $g_{J}$ is the Land\'{e} $g$\,-\,factor, $E_{n}$ and $\mid\!n \rangle$ are the $n$th eigenvalue and eigenfunction, respectively.  $J_{i}$ ($i$\,=\,$x$, $y$ and $z$) is a component of the angular momentum,  and
$\Delta_{m,n}\,=\,E_{n}\, - \,E_{m}$, $Z\,=\,\sum_{n}e^{-\beta
E_{n}}$ and $\beta\,=\,1/k_{\rm B}T$.  The magnetic susceptibility
including the molecular field contribution $\lambda_{i}$ is given
by
\begin{equation}
\label{eqn4}
\chi^{-1}_{i} = \chi_{{\rm CEF}i}^{-1} - \lambda_{i}.
\end{equation}

\begin{figure}[t]
\includegraphics[width=0.35\textwidth]{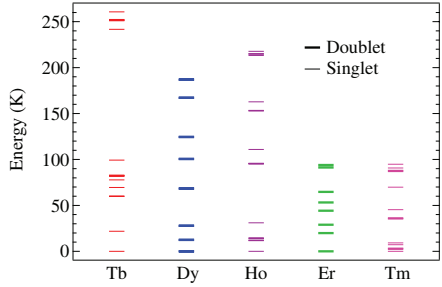}
\caption{\label{fig19} (Color online) CEF energy level splitting of the ground state of the R$^{3+}$ ions in R$_2$CoGa$_8$ compounds.}
\end{figure}

The inverse susceptibility of the R$_2$CoGa$_8$ (R = Tb, Dy, Ho, Er and Tm) as shown in Fig.~[\ref{fig7}(b), \ref{fig9}(b), \ref{fig11}(b), \ref{fig14}(b) and \ref{fig16}(b)] respectively was fitted to the above discussed CEF model Eqn.~(\ref{eqn2}-\ref{eqn4}).  The values of the CEF parameters thus obtained are presented in Table~\ref{table4} and the corresponding energy levels are shown in Fig.~\ref{fig19}.  The
\begin{table*}[!]
\centering
\begin{ruledtabular}
\begin{tabular}{cccccccccc}
R   &  $B_{2}^{0}$ & $B_{4}^{0}$ &$B_{4}^{4}$ & $B_{6}^{0}$ & $B_{6}^{4}$ & $\lambda^{[100]}$ & $\lambda^{[001]}$ & $\mathcal{J}_{\rm ex}^{[100]}/k_{\rm B}$  & $\mathcal{J}_{\rm ex}^{[001]}/k_{\rm B}$ \\
&  (K)  &  (K)  & (K)  & (K)   & (K)  & (mol/emu)   & (mol/emu) & (K) & (K)\\
\hline \\
Tb   &   -1.61  &  0.0049    &   -0.0515   & -7.0~$\times~10^{-5}$ & 1.63~$\times~10^{-3}$   &   -2.95  &   -4.7  & -2.24   &   -4.94 \\

Dy   &   -0.7  &  0.005    &  0.001   & 9.0~$\times~10^{-8}$ & 1.0~$\times~10^{-4}$   &   -2.5  &   -2.0  &  -1.28   &   -1.94 \\

Ho   &   -0.22  &  -1.45~$\times~10^{-3}$    &   -0.0314   & -2.0~$\times~10^{-5}$ & -2.0~$\times~10^{-5}$   &   -1.2  &   -1.2 &  -0.51   &   -0.59\\

Er   &   0.089  &  1.3~$\times~10^{-4}$    &   0   & 4.3~$\times~10^{-6}$ & 2.5~$\times~10^{-4}$   &   -0.782  &   -0.274  &   -0.35   &   -0.32\\

Tm   &   0.35  &  1.72~$\times~10^{-6}$    &   0.0248   & -8.46~$\times~10^{-6}$ & 9.1~$\times~10^{-4}$   &   0.723  &   0  &   -0.53   &   -0.075 \\

\end{tabular}
\end{ruledtabular}
\caption{CEF paramters for R$_2$CoGa$_8$ compounds obtained from the inverse susceptibility fit}
\label{table4}
\end{table*}
dominant crystal field parameter $B_{2}^{0}$ is negative for Tb, Dy and Ho compounds. This is consistent with the uniaxial ([001] as an easy axis) magnetic anisotropy present in these compounds. The sign of the $B_{2}^{0}$ changes for Er and Tm compounds which is consistent with the change in the easy axis of magnetization. For Er$_2$CoGa$_8$, the estimated $B_{2}^{0}$ is 0.089~K, which is close to zero, indicating that the compound is on the border line of the magnetic anisotropy crossover. The current set of  CEF parameters could even explain the crossover in the magnetic susceptibility of Er$_2$CoGa$_8$. According to mean field theory, the CEF parameter $B_{2}^{0}$ can be related to the exchange constant and paramagnetic Curie temperature by the relation~\cite{Jensen}

\begin{equation}
\label{eqn5}
\theta_{\rm p}^{[001]} = \frac{J(J+1)}{3k_{\rm B}} \mathcal{J_{\rm ex}}^{[001]} - \frac{(2J-1)(2J+3)}{5k_{\rm B}} B_{2}^{0} ,
\end{equation}
\begin{equation}
\label{eqn6}
\theta_{\rm p}^{[100]} = \frac{J(J+1)}{3k_{\rm B}} \mathcal{J_{\rm ex}}^{[100]} + \frac{(2J-1)(2J+3)}{10k_{\rm B}} B_{2}^{0}.
\end{equation}
 
The obtained value for $\mathcal{J}_{\rm ex}$ along both the crystallographic directions is presented in Table~\ref{table4}. The negative value along both the principal directions implies that the anitferromagnetic interaction is dominant in this series of compounds. In the case of R$_2$CoGa$_8$ (R = Tb, Dy and Ho) the exchange constant $\mathcal{J}_{\rm ex}$, has a higher value along [001] compared to that of [100] direction.  This implies that the antiferromagnetic exchange interaction is stronger along the [001] direction, thus supporting our experimental observation of magnetic easy axis for these compounds.  Whereas, for R$_2$CoGa$_8$ (R = Er and Tm) the  $\mathcal{J}_{\rm ex}$ is dominant along the [100] direction, which is again consistent with our magnetic susceptibility data. 

The magnetic transition temperature ($T_{\rm M}$) of a compound in presence of crystalline electric field is given by Noakes~\textit{et al}~\cite{Noakes} as,

\begin{equation}
\label{eqn7}
T_{\rm M} = 2 \mathcal{J}_{\rm ex} (g_J - 1)^2 \langle J_{z}^{2} (T_{\rm M}) \rangle_{\rm CEF},
\end{equation}
	 
Where $\langle J_{z}^{2}(T_{\rm M})\rangle_{\rm CEF}$ is the expectation value of $J_{\rm z}^2$ at $T_{\rm M}$ under the influence of crystalline electric field alone.  $\mathcal{J}_{\rm ex}$ is the exchange constant for the RKKY exchange interaction between the rare earth atoms. Now for the compounds with tetragonal structure the dominating crystal field term is  $B_{2}^{0}$, so neglecting the higher order term and when the ordering is along the [001] direction, the above equation  can be rewritten as~\cite{Noakes}

\begin{equation}
\label{eqn8}
T_{\rm M} = 2 \mathcal{J}_{\rm ex} (g_J - 1)^2 \sum_{J_z} \frac {J_{z}^{2} {\rm exp}({-3 J_{z}^{2}B_{2}^{0}}/ {T_{\rm M}})} { {{\rm exp}(-3 J_{z}^{2}B_{2}^{0}/T_{\rm M})}}.
\end{equation}

The transition temperature T$_{\rm M}$ was calculated using the values of the exchange constant $\mathcal{J_{\rm ex}}$ obtained from Eqn.~\ref{eqn5} for the easy-axis of magnetization and with the corresponding $B_{2}^{0}$ crystal field parameter. The magnetic transition temperature for Tb$_2$CoGa$_8$ and Dy$_2$CoGa$_8$ thus estimated was found to be ($\approx$ 82~K) and ($\approx$ 22~K), respectively.  This shows the enhancement in the transition temperature in presence of CEF compared to that expected from de-Gennes scaling. Though the estimated $T_{\rm N}$ in  Tb$_2$CoGa$_8$ is unexpectedly high, the preceding analysis shows that the enhancement of $T_{\rm N}$ of  Tb$_2$CoGa$_8$ can be due to CEF effects.

\section{Conclusion}

To conclude, we have successfully grown the single crystals of R$_2$CoGa$_8$ (R = Gd, Tb, Dy, Ho, Er, Tm, Lu and Y) for the first time by using Ga as flux. This series of compounds forms only with the higher rare earths. The phase purity of the crystals was confirmed by means of powder X-ray diffraction. Y$_2$CoGa$_8$ and Lu$_2$CoGa$_8$ show diamagnetic behavior indicating a low density of states at the Fermi level and a  filled $3d$ of Co band.  Compounds with magnetic rare earths order antiferromagnetically at low temperatures. The N\'{e}el temperature of Tb$_2$CoGa$_8$ and Dy$_2$CoGa$_8$ deviates appreciably from expected value of de-Gennes scaling. The reason is attributed to crystal field effects. The easy axis of magnetization for R$_2$CoGa$_8$ (R = Tb, Dy and Ho) is along the crystallographic [001] direction. It changes to (100) plane for Er$_2$CoGa$_8$ and Tm$_2$CoGa$_8$.


\begin{thebibliography}{99}

\bibitem{Kalychak}Kalychak Ya. M, Zeremba V. I, Baranyak V. M, Bruskov V. A and Zavalij P. Yu, Izv. Acad, Nauk SSSR Metally {\bf 1}, 209 (1989).

\bibitem{Thompson}J.D. Thompson, R. Movshovich, Z. Fisk, F. Bouquet, N.J. Curro, R.A. Fisher, P.C. Hammel, H. Hegger, M.F. Hundley, M. Jaime, P.G. Pagliuso, C. Petrovic, N.E. Phillips and J.L. Sarrao, J. Magn. Magn. Mater. {\bf 226-230}, 5 (2001).

\bibitem{Nicklas}M. Nicklas, V. A. Shidorov, H. A. Borges, P. G. Pagliuso, C. Petrovik, Z. Fisk, J.L. Sarrao and J.D. Thompson, Phys. Rev. B {\bf 67}, 020506 (2003).

\bibitem{Chen}G. Chen, S. Ohara, M. Hedo, Y. Uwatoko, K. Saito, M. Sorai and I. Sakamoto, J. Phys. Soc. Japan {\bf 71} 2836 (2002).

\bibitem{Devang}Devang A. Joshi, C.V. Tomy and S.K. Malik, J. Phys.: Condens. Matter {\bf 19} 136216 (2007).

\bibitem{Cirafici}S. Cirafici and E. Franceschi, J. Less-Common Met. {\bf 77}, 269 (1981).

\bibitem{Kalychak2}Ya. M. Kalychak, J. Alloys Compounds 291 (1999) 80

\bibitem{Buschow}K. H. J. Buschow, J. Chem. Phys., {\bf 50}, 137 (1969)

\bibitem{Dunlap}B. D. Dunlap, L. N. Hall, F. Behroozi, G. W. Crabtree and D. G. Niarchos, Phys. Rev. B {\bf 29} 6244 (1984).

\bibitem{Noakes}D. R. Noakes and G. K. Shenoy, Phys. Lett. {\bf 91A}, 35 (1982).

\bibitem{Mallik}R. Mallik and E. V. Sampathkumaran, Phys. Rev. B {\bf 58}, 9178 (1998).

\bibitem{Rotter}M. Rotter, M. Loewenhaupt, M. Doerr, A. Lindbaum and H. Michor, Phys. Rev. B {\bf 64} 014402 (1998).

\bibitem{Myers}K. D. Myers, S. L. Budko, I. R. Fisher, Z. Islam, H. Kleinke, A. H. Lacerda and P. C. Canfield, J. Magn. Magn. Mater. {\bf 205}, 27 (1999).

\bibitem{Hieu}N. V. Hieu, T. Takeuchi, H, shishido, C. Tonohiro, T. Yamada, H. Nakashima, K. Sugiyama, R. Settai, T. D. Matsuda, Y. Haga, M. Hagiwara, K. Kindo, S. Araki, Y. Nozue and Y. Onuki, J. Phys. Soc. Japan {\bf 76}, 064702 (2007).

\bibitem{Frontzek}M. Frontzek, A. Kreyssig, M. Doerr, M. Rotter, G. Behr, W. L\"{o}ser, I. Mazilu and M. Loewenhaupt, J. Magn. Magn. Mater., {\bf 301}, 398 (2006).

\bibitem{Stevens} K. W. H. Stevens, Proc. Phys. Soc., London, Sect. A{\bf 65}, 209 (1952).

\bibitem{Hutchings} M. T. Hutchings, in {\it Solid State Physics: Advances in Research and Applications}, edited by F. Seitz and B. Turnbull (Academic, New York, 1965), Vol.16, p.227.

\bibitem{Jensen}J. Jensen and A. R. Mackintosh, Rare earth magnetism structures and excitations, Calrendon press, Oxford {\bf Chap 2}, p. 73 (1991).


\end{thebibliography}
\end{document}